\documentclass[preprint]{aastex62} 
\usepackage{amsmath}
\usepackage{comment}
\usepackage{color}

\newcommand{\g}{{\rm g}}

\newcommand{\cm}{{\rm cm}}

\newcommand{\AU}{{\rm AU}}

\newcommand{\K}{\rm K}
\newcommand{\yr}{\rm yr}

\newcommand{\Msun}{M_{\odot}}

\begin{document}

\title{Enrichment of Jupiter's atmosphere by late planetesimal bombardment}

\correspondingauthor{Sho Shibata}
\email{sshiba@uzh.ch}

\author{Sho Shibata}
\affiliation{Institute for Computational Science (ICS), University of Zurich}

\author{Ravit Helled}
\affiliation{Institute for Computational Science (ICS), University of Zurich}


  \begin{abstract}
    Jupiter's atmosphere is enriched with heavy elements by a factor of about 3 compared to proto-solar. The origin of this enrichment and whether it represent the bulk composition of the planetary envelope remain unknown. 
    Internal structure models of Jupiter suggest that its envelope is separated from the deep interior and that the planet is not fully mixed. This implies that Jupiter's atmosphere was enriched with heavy elements just before the end of its formation. Such enrichment can be a result of late planetesimal accretion. 
    However, in-situ Jupiter formation models suggest the decreasing accretion rate with increasing planetary mass, which cannot explain Jupiter's atmospheric enrichment.
    In this study, we model Jupiter's formation and show that an migration of proto-Jupiter from $\sim$ 20 AU to its current location can lead to a late planetesimal accretion and atmospheric enrichment.
    Late planetesimal accretion does not occur if proto-Jupiter migrates only a few AU. 
    We suggest that if Jupiter's outermost layer is fully-mixed and is relatively thin (up to $\sim$ 20\% of its mass), such late accretion can explain its measured atmospheric composition. 
    It is therefore possible that Jupiter underwent significant orbital migration followed by late planetesimal accretion.
   \end{abstract}




\section{Introduction}
\label{sec:introduction}
    The Galileo probe measured the elemental abundances in Jupiter's atmosphere and found that several heavy elements are enriched by a factor of $\sim$ 3 compared to a proto-solar composition \citep[e.g.,][]{Owen+1999,Wong+2004,Atreya+2020}. 
    Also the recent measurement of Jupiter's water abundance by Juno imply  that oxygen is enriched by a factor of a few \citep{Li+2020}.
    
    The origin of the heavy-element enrichment of Jupiter's atmosphere remains unknown and several ideas have been suggested to explain this enrichment. 
    One idea is that the atmospheric enrichment is caused by the erosion of a primordial heavy-element core \citep[e.g.,][]{Stevenson+1982,Guillot+2004,Oberg+2019,Bosman+2019}. 
    However, in this case the materials dissolved into the deep interior must be mixed by convection and be delivered to the upper envelope. 
    Since recent structure models of Jupiter imply that the planet is not fully convective \citep[e.g.,][]{Leconte+2013, Wahl+2017,Vazan+2018,Debras+2019}, the validity of this explanation is questionable and should be investigated in detail. 
    Alternatively, Jupiter's atmospheric enrichment could be a result of the accretion of enriched disk gas \citep{Guillot+2006,Bosman+2019,Schneider+2021}.
    However, this scenario  cannot reproduce the atmospheric enrichment of water and refractory materials \citep{Schneider+2021}.
    Finally, it is possible that Jupiter's atmosphere has been enriched by late accretion of 
    \replaced{heavies}{heavy elements}, 
    in the form of planetesimal accretion, as we explore in this work.  
    
    Previous investigations of Jupiter's formation considered planetesimal accretion in the context of in-situ formation where the proto-Jupiter grows at 5.2 AU \citep[e.g.,][]{Zhou+2007,Shiraishi+2008,Shibata+2019,Venturini+2020,Podolak+2020}.
    In this case, as the gas accretion rate increases, the planetesimal accretion rate decreases. 
    Therefore, if Jupiter is not fully convective the enrichment of its outer envelope is difficult to explain 
    \added{because the accreted planetesimals are mainly deposited in the deep interior  and not delivered to the upper envelope}.
    However, there is a clear theoretical indication that planets migrate \citep[e.g.][]{Bitsch+2015,Kanagawa+2018,Ida+2018,Bitsch+2019a,Tanaka+2020}.
    During the planetary migration, planetesimals can be captured
    \added{by the planet}
    \citep[e.g.][]{Alibert+2005}, and it was recently shown that the migration rate regulates the planetesimal accretion rate \citep{Shibata+2020,Shibata+2021,Turrini+2021}.
    It was shown in \citet{Shibata+2021} that rapid planetesimal accretion occurs in the limited region which we refer to as the "sweet spot for planetesimal accretion" (SSP).
    The SSP is located around $\lesssim10\AU$ for planets smaller than Jupiter, suggesting that proto-Jupiter enters the SSP after a large fraction of its envelope has already accumulated.
    If this is the case, a non-negligible amount of planetesimals can be accreted into the outer layer of the proto-Jupiter and lead to an enrichment of its atmosphere. 
    
    In this letter, we simulate Jupiter's formation including planetary migration and investigate the accretion rate of planetesimals.
    In Sec.~\ref{sec:method}, we describe our numerical model and the formation pathways of proto-Jupiter we consider.
    Our results are presented in Sec.~\ref{sec:results} where we show that rapid planetesimal accretion occurs just before the end of Jupiter formation.
    A discussion on the connection to Jupiter's measured atmospheric metallicity is  
    presented in Sec.~\ref{sec:discussion}.
    Finally, our conclusions are discussed in Sec.~\ref{sec:conclusion}.

\section{Methods}
\label{sec:method}
    We perform orbital integration calculations  of planetesimals around a proto-planet growing via disk gas accretion (increasing planetary mass $M_{\rm p}$) and migrating inward due to the tidal interaction with the surrounding gaseous disk (decreasing planetary semi-major axis $a_{\rm p}$).
    Our simulations begin from the rapid gas accretion phase for given planetary mass  $M_{\rm p,0}$ and semi-major axis $a_{\rm p,0}$.
    We assume that there are many single-sized planetesimals with a radius of $R_{\rm pl}$ around the protoplanet's orbit.
    The protoplanet then encounters these planetesimals and can capture some of them. 
    Planetesimals are represented by test particles and are therefore only affected by the gravitational forces from the central star (with a mass of $M_{\rm s}=M_{\odot}$) and the protoplanet, as well as the drag force of the gaseous disk. 
    To model the drag force we follow the model of  \citet{Adachi+1976}.
    The dynamical integration for the bodies is performed using the numerical framework presented in \citet{Shibata+2019}.
    
\begin{figure}
      \begin{center}
        \includegraphics{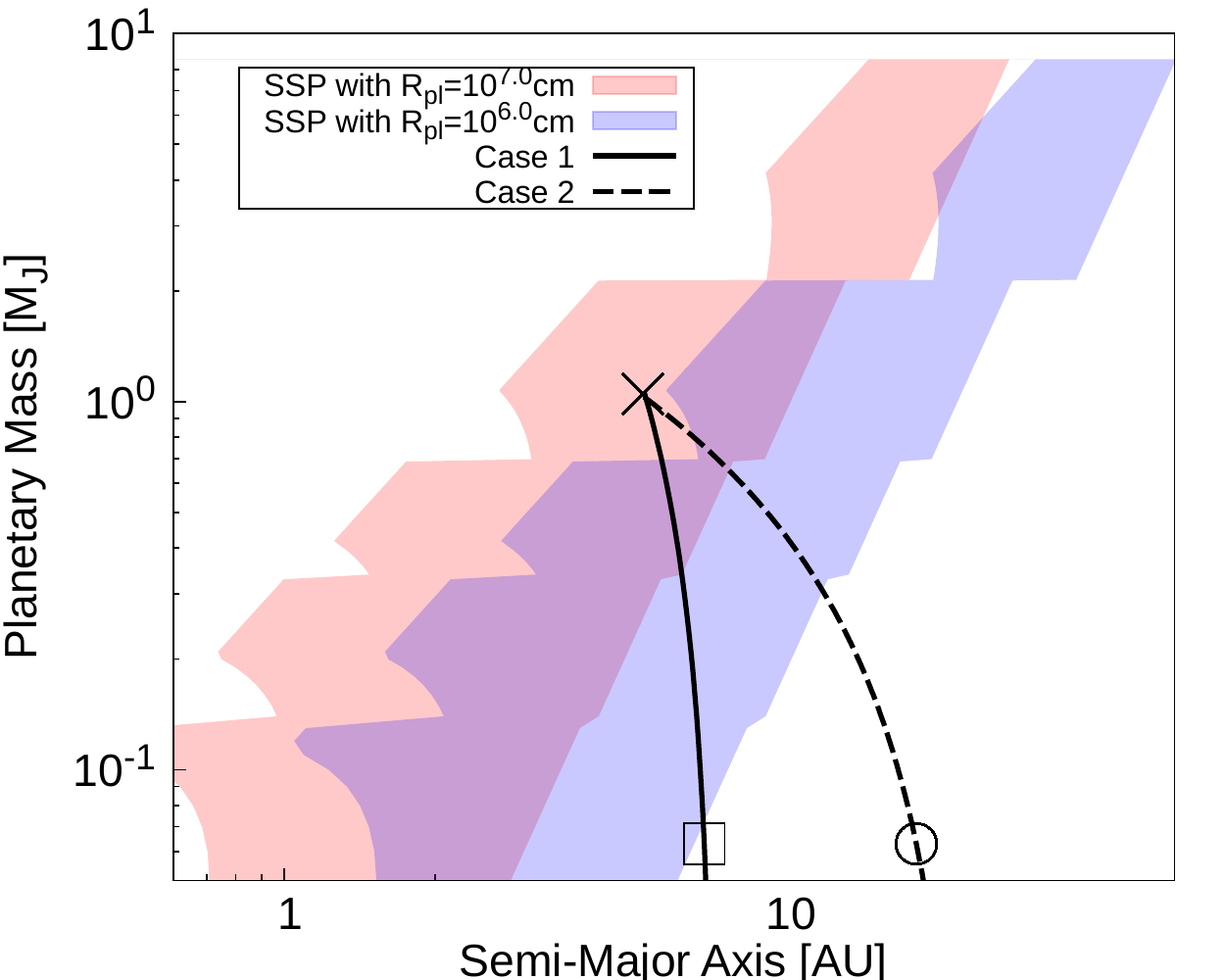} 
        \caption{
        Orbital evolution pathways used in our model.
        Solid and dashed lines show the evolution pathways in Case 1 and Case 2, respectively.
        Square and circle points are core formation location for Case 1 and Case 2, respectively.
        Cross point corresponds to Jupiter.
        The red and blue areas show the sweet spot for planetesimal accretion (SSP) \citep{Shibata+2021} with $R_{\rm pl}=10^{7.0}\cm$ and $R_{\rm pl}=10^{6.0}\cm$, respectively.
        }
        \label{fig:ap_mp}
      \end{center}
    \end{figure}

    We adapt the formation model of  \citet{Tanaka+2020}, where both the gas accretion timescale $\tau_{\rm acc}$ and planetary migration timescale $\tau_{\rm tide}$ depend on the gap structure opened by the protoplanet's tidal torque.
    In this model, the effect of the density profile of the disk is cancelled, and the relation between the two timescales is given by: 
    \begin{align}
        \frac{\tau_{\rm tide}}{\tau_{\rm acc}} = \left|\frac{d \ln{M_{\rm p}}}{d \ln{a_{\rm p}}}\right|  = \left( \frac{M_{\rm p}}{M_{\rm th}} \right)^{-2/3}, \label{eq:timescale_fraction}
    \end{align}
    where $M_{\rm th}$ is the threshold mass determined by the gas accretion and migration models.
    In \citet{Tanaka+2020}, $M_{\rm th}$ is estimated as $\sim10^{-2}$ \added{$M_{\rm s}$}.
    Their gas accretion model assumed that most of the disk gas entering the hill sphere is accreted by the planet. 
    However, as pointed in \citet{Ida+2018}, recent hydrodynamic simulations clearly show that this is an over estimate of the gas accretion rate  \citep[e.g.,][]{Szulagyi+2016,Kurokawa+2018}.
    We therefore consider two formation pathways with $M_{\rm th}=10^{-2} M_{\rm s}$ (Case 1) and $M_{\rm th}=10^{-3} M_{\rm s}$ (Case 2). 
    Figure~\ref{fig:ap_mp} shows the formation pathways of proto-Jupiter for these two cases where the solid and dashed lines correspond to Case 1 and Case 2, respectively.
    We set $M_{\rm p,0}=6\times10^{-5} M_{\odot}\sim20M_{\oplus}$ and adjust $a_{\rm p,0}$ as $6.9\AU$ for Case 1 (square point) and $18.3\AU$ for Case 2 (circle point) in order to ensure that the protoplanet reaches Jupiter's mass at its current location (cross point in Fig.~1). 
    It is clear that other formation paths are possible, however, in this study we focus only on two 
    \replaced{end-member}{typical}
    cases representing short and long migration in order to investigate how they compare in terms of late heavy-element enrichment. 
    
    The SSP 
    \added{
    is determined by the planetary migration rate and the damping rate of the  plaentesimal's orbit due to disk gas drag.
    Therefore, the location of the SSP
    }
    depends on various parameters, such as the disk viscosity, aspect ratio, and planetesimal size \citep{Shibata+2021}.
    In order to investigate the effect of the SSP, we consider different planetesimal sizes $R_{\rm pl}$. 
    In Fig.~\ref{fig:ap_mp}, we plot the formation pathways of the proto-planet and the SSP when considering two different planetesimal sizes.
    As we show below the relative position between the formation pathways and the SSP affects the enrichment of the planetary envelope.
    The detailed model and other parameters used in our simulations are presented in the Appendix \ref{app:model}.

\section{Results}
\label{sec:results}
    \begin{figure*}
      \begin{center}
        \includegraphics[width=160mm]{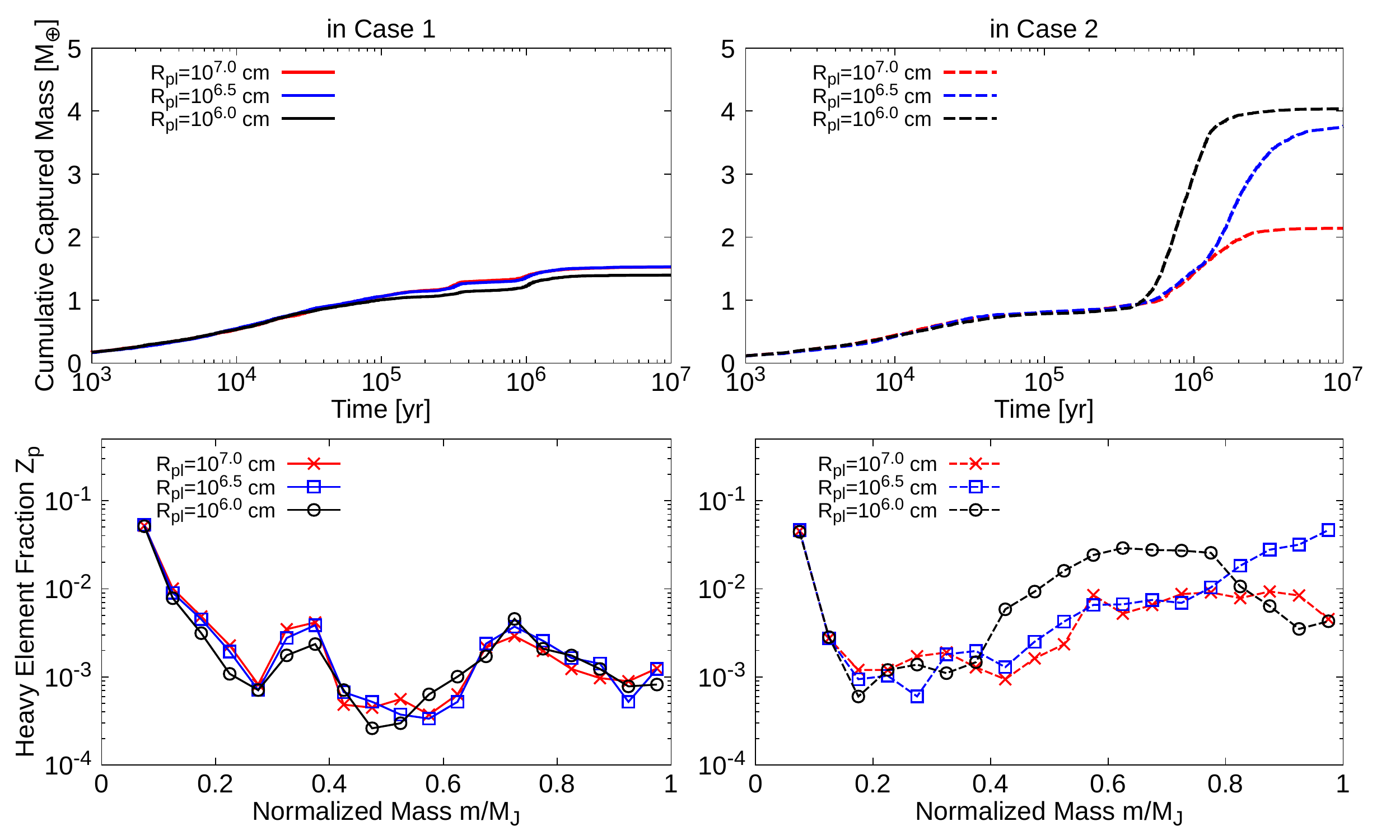} 
        \caption{
        {\bf Upper panels:} The cumulative captured mass of planetesimals as a function of calculation time $t-t_0$ in Case 1 (left) and in Case 2 (right).
        The different colours correspond to the different size of planetesimals. 
        The planetesimal accretion rate decreases with time in Case 1, however, second rapid planetesimal accretion occurs before the end of its formation in Case 2.
        {\bf Lower panels:} The heavy-element mass fraction $Z_{\rm p}$ formed in Jupiter's envelope as a function of the normalised mass $m/M_{\rm p}$.
        \added{
        The bin's width  is set as $0.05~ m/M_{\rm J}$.
        }
        Planetesimals accreted during the second accretion phase are deposited into the outer envelope.
        }
        \label{fig:Result}
      \end{center}
    \end{figure*}
    
    Figure~\ref{fig:Result} shows the results of our simulations.
    The upper panels present the cumulative captured mass of planetesimals $M_{\rm cap}$ as a function of calculation time $t-t_0$. 
    In Case 1 (left panel), the cumulative captured mass gradually increases with time although the accretion rate decreases.
    This is because the expanding speed of the feeding zone decreases with the increase of gas accretion timescale and planetary mass \citep{Shibata+2019}. 
    This results in the depletion of planetesimals inside the feeding zone.
    For Case 2 (right panel), the accretion rate decreases until $t-t_0\lesssim10^6\yr$.
    Planetesimal accretion nearly stops when  $t-t_0\sim10^5\yr$.
    However, a second planetesimal accretion phase occurs before the end of Jupiter's formation. 
    As shown in Fig.~\ref{fig:ap_mp}, proto-Jupiter enters the SSP before the end of its formation, which triggers the second phase planetesimal accretion.
    Before it enters the SSP, many planetesimals are shepherded by the mean motion resonances.
    This leads to a large amount of planetesimals that enters the feeding zone when proto-Jupiter reaches the SSP.
    
    The planetesimal accretion rate increases with decreasing $R_{\rm pl}$. 
    This is because the SSP moves outward with decreasing $R_{\rm pl}$ and the length of the evolutionary pathway that overlaps with the SSP is longer for smaller planetesimals.
    On the other hand, the initiation of the second phase of planetesimal accretion occurs earlier for smaller planetesimals because proto-Jupiter enters the SSP earlier.
    
    The lower panels of Fig.~\ref{fig:Result} show the heavy-element distribution formed in Jupiter envelope $Z_{\rm p}$ by the end of the simulations.
    To estimate $Z_{\rm p}$, we 
    \added{
    assume that the captured planetesimals are deposited in the outer regions at that time and
    }
    neglect any mixing processes; namely $Z_{\rm p}$ is obtained from the planetesimal accretion rate normalised by the mass growth rate $\dot{M}_{\rm cap}/\dot{M}_{\rm p}$.
    For Case 1 (left panel), the planetesimal accretion rate decreases with time and the outer envelope is barely enriched with planetesimals.
    On the other hand, for Case 2 (right panel), the planetesimals accreted during the  second accretion phase are deposited in the outer envelope ($\gtrsim 0.5 M_{\rm J}$).
    This is most profound  for $R_{\rm pl}=10^{6.5}$, where the metallicity of Jupiter's atmosphere  can be enhanced by a factor of a few. 
    It should be noted, however, that the final metallicity of Jupiter's atmosphere would depend on the mixing and settling of the heavy elements accreted at this late stage. 
    We discuss this topic in more detail in Sec.~\ref{sec:discussion}.
    
    It should be noted that in this study we focus on the enrichment of Jupiter's atmosphere.
    However, formation models should also reproduce the total heavy-element mass in the planet. 
    Interior models of Jupiter that fit Juno gravity data suggest that Jupiter's interior is enriched with a few tens M$_{\oplus}$ of heavy elements although the exact heavy-element mass is not well-constrained. 
    In this study, we begin the simulation of proto-Jupiter with a mass of $\sim20M_{\oplus}$ of heavy elements without specifying the formation process of the heavy-element core.
    Core formation via planetesimal/pebble accretion is expected to lead to deep interiors that are heavy-element dominated in composition and for the build-up of composition gradients  \citep[e.g.][]{Lozovsky+2017,Helled+2017,Valletta+2020}.
    We therefore assume that the deep interior of the forming planet consists of mostly heavy elements. 

\section{Discussion}
\label{sec:discussion}
    \subsection{Enrichment of Jupiter's atmosphere}
    \begin{figure}
      \begin{center}
        \includegraphics{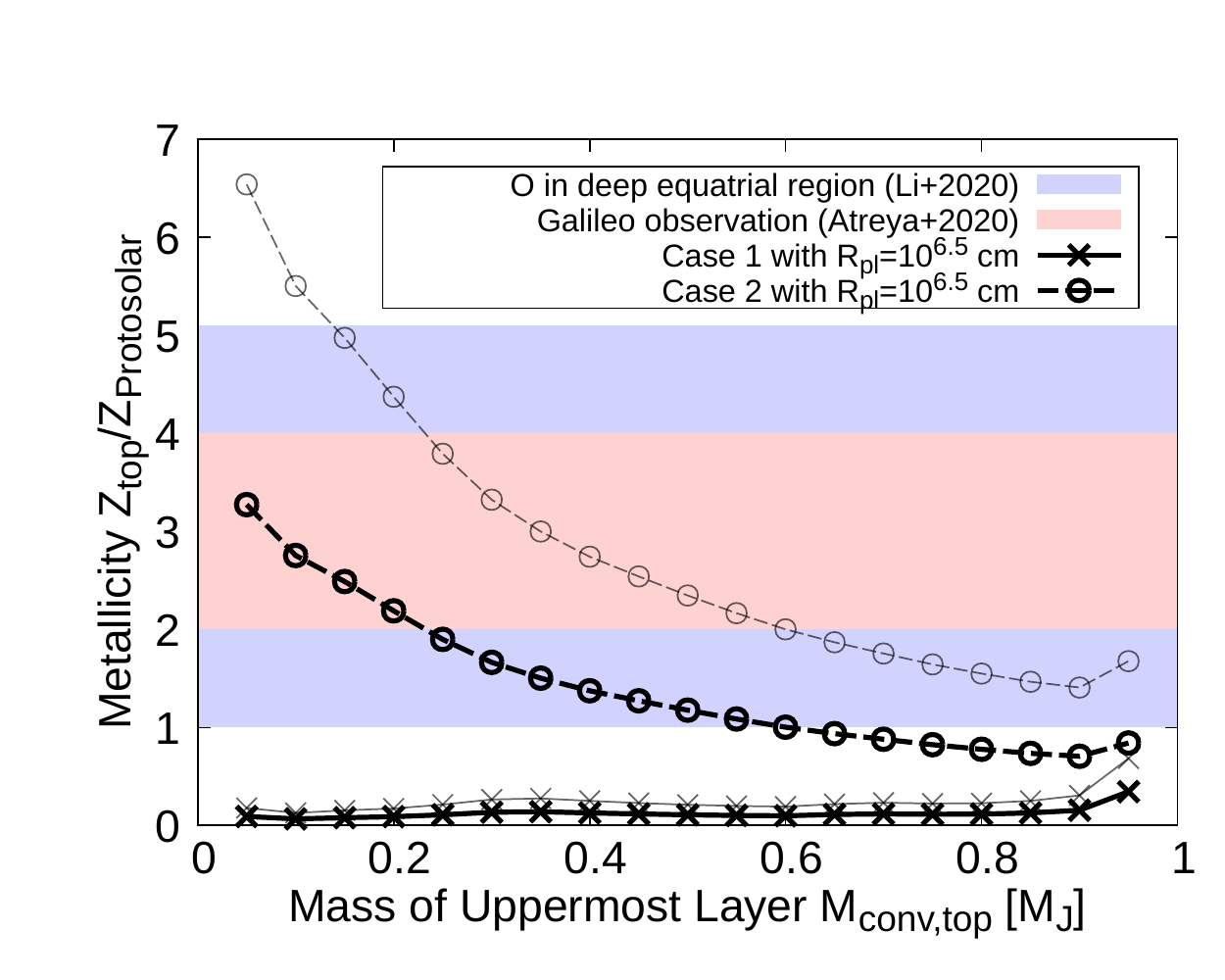}
        \caption{
        Metallicity of uppermost layer $Z_{\rm top}$ as a function of the mass of uppermost layer $M_{\rm conv,top}$. 
        The metallicity is normalised by the proto-solar metallicity $Z_{\rm protosolar}$.
        We adapt 0.0142 obtained in \citet{Asplund+2009} to be consistent with our disk model.
        The blue and red areas show the oxygen abundance retrieved by Juno ($2.7^{+2.4}_{-1.7}$, \citet{Li+2020}) and atmospheric composition suggested by Galileo ($2-4$, \citet{Atreya+2020}).
        Solid and dashed lines show the results in Case 1 and Case 2, respectively.
        Here, we show the cases of $R_{\rm pl}=10^{6.5} \cm$.
        We also plot the cases where planetesimal disk is twice heavier than the disk in our model with thin lines.
        }
        \label{fig:Mixing_Z}
      \end{center}
    \end{figure}
    
    The accreted heavy elements can be redistributed by the mixing processes in Jupiter's envelope that can occur over a timescale of Gyrs.
    Here for simplicity, we assume that the uppermost layer of Jupiter's envelope with a mass $M_{\rm conv,top}$ is  separated from the deeper interior and that the accreted heavy-element mass deposited in this region $M_{\rm cap,top}$ is uniformly distributed. In other words, we assume that the outermost region of Jupiter's envelope is convective and homogeneously mixed and that the envelope below this layer has a different composition. This assumption is in fact consistent with recent models of Jupiter's interior \citep[e.g.,][]{Vazan+2018}. 
    In this case, the metallicity of Jupiter's atmosphere (outer envelope) $Z_{\rm top}$ is given  by: \begin{align}
        Z_{\rm top} = \frac{M_{\rm cap,top}}{M_{\rm conv,top}}.
    \end{align}
    
    Figure~\ref{fig:Mixing_Z} shows $Z_{\rm top}$ as a function of $M_{\rm conv,top}$.
    The solid and dashed black lines correspond to Case 1 and Case 2, respectively. 
    Here, we show the cases of $R_{\rm pl}=10^{6.5}\cm$ with which the most efficient atmospheric enrichment is achieved.
    The thin lines show the results for the cases where we use a planetesimal disk that is twice more massive than our baseline model.
    In Case 1, $Z_{\rm top}$ is significantly lower than proto-solar metallicity independently of $M_{\rm conv,top}$. 
    This is because the total mass of accreted planetesimals is smaller than $1M_{\oplus}$ and the planetesimals are accreted at early stages and are deposited in the deep interior. 
    Even when we consider a planetesimal disk that is several times more massive, it is difficult to explain Jupiter's enriched atmosphere in Case 2.
    
    On the other hand, for Case 2, several M$_{\oplus}$ of heavy elements are accreted during the late phases of Jupiter's formation leading to the enrichment of the planetary uppermost envelope as we show in  Fig.~\ref{fig:Mixing_Z}. 
    We find that in this case $Z_{\rm top}$ is enhanced in comparison to proto-solar metallicity for $M_{\rm conv,top}\lesssim0.6M_{\rm J}$.
    The blue and red areas correspond to the measured water abundance by Juno \citep{Li+2020} and the elemental abundance measured by the Galileo probe \citep{Atreya+2020}, respectively. 
    We find that Jupiter's atmospheric metallicity can be explained with $M_{\rm conv,top}\lesssim0.2M_{\rm J}$.
    The maximum required value of $M_{\rm conv,top}$ can increase when considering larger sizes of the planetesimal disk. 
    It should be noted, however, that the 
    \replaced{depth}{size}
    of the uppermost convective envelope of Jupiter is not well-constrained and changes with different structure models \citep[e.g.,][]{Wahl+2017,Vazan+2018,Debras+2019}. 
    Using a planetary evolution model, \citet{Vazan+2018} found that a primordial composition gradient in the deep interior can be partially eroded by convective mixing leading to a large convective envelope ($60\%$ of Jupiter mass) for Jupiter today. 
    In this case, a more massive  planetesimal disk 
    \added{as well as smaller planetesimals and/or larger planetary capture radius could increase the heavy-element accretion efficiency (see Appendix~\ref{app:model})}
    are required to reproduce the measured elemental abundances in Jupiter's atmosphere.
    To explain Jupiter's atmospheric enrichment, our results clearly favour a small outer convective layer for Jupiter as proposed by \citet{Debras+2019} ($M_{\rm conv,top}\sim0.1M_{\rm J}$).
    
    
    \subsection{Very volatile materials}\label{sec:Dis_VM}
    \begin{figure}
      \begin{center}
        \includegraphics{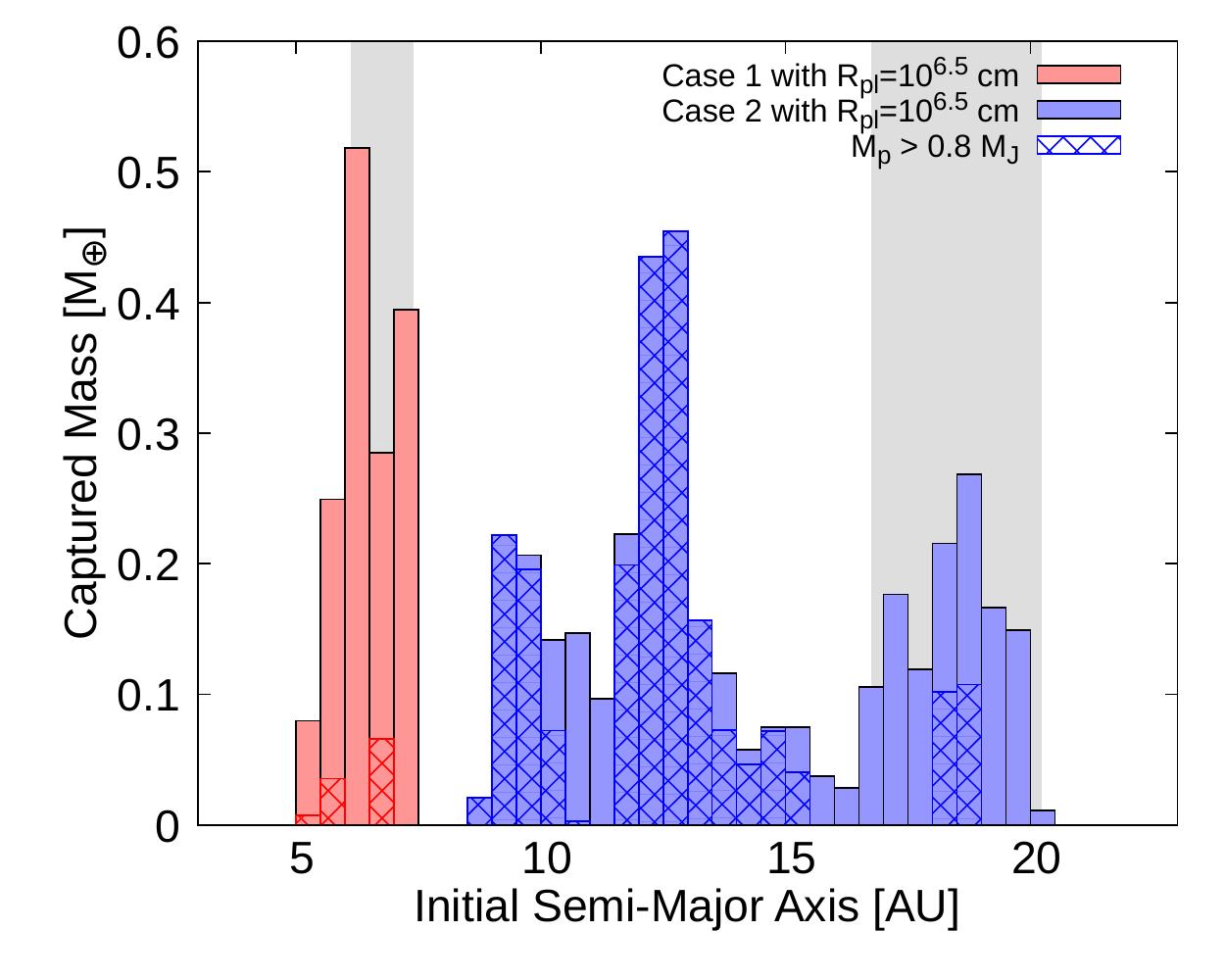}
        \caption{
        Mass of captured planetesimals as a function of the initial semi-major axis of planetesimals.
        The red and blue correspond to Case 1 and Case 2, respectively.
        The meshed textures show the mass of captured planetesimals when $M_{\rm p}>0.8M_{\rm J}$.
        Here we show the cases of $R_{\rm pl}=10^{6.5}\cm$.
        Bin-width is set as $0.5 \AU$.
        The gray filled areas show the initial feeding zones of proto-Jupiter.
        }
        \label{fig:Result_axi0_fcap}
      \end{center}
    \end{figure}
    
    The measurement of the Galileo probe also finds that very volatile materials are also enriched in Jupiter's atmosphere. 
    Such elements are expected to condense in a cold environment where the temperature is $\lesssim30\K$ \citep{Atreya+2020}.
    Figure~\ref{fig:Result_axi0_fcap} shows the accreted heavy-element mass as a function of the initial semi-major axis of the planetesimals.
    The mass of accreted planetesimals when $M_{\rm p}>0.8M_{\rm J}$ is indicated with meshed textures.
    In Case 2, the planetesimals accreted onto the upper envelope mainly come from the relatively cold outer region of the disk ($10\sim15\AU$).
    However, even for the case of an optically thick disk \citep{Sasselov+2000}, the mid-plane temperature is too high for very volatile materials, such as N$_2$ and Ar, to condense into planetesimals.
    \added{
    It is also possible that Jupiter migrated from a larger radial distance, such as $>30 \AU$  \citep[e.g.][]{Bitsch+2015,Bitsch+2019a}.
    However, the migration of a core from several tens  of $\AU$ to $5 \AU$ requires rapid formation of the core at these large distances, and it is still unclear whether a  core can form in such outer disk in timescales that are significantly shorter than the disk's lifetime. 
    Additional processes could help in  enriching} the atmosphere, for example, by decreasing the \added{local} mid-plane temperature with a shadow of the inner disk \citep[e.g.][]{Ohno+2021}, or by increasing the metallicity of the accreting disk gas towards the end of disk depletion \citep{Guillot+2006b}.
    \added{
    Also, volatiles must keep being condenced during the accretion process which might be unrealistic given the expected heating by proto-Jupiter's luminosity  \citep{Barnett+2022} or ablation by the disk gas drag \citep[e.g.][]{Eriksson+2021}.
    The accretion of volatile materials by Jupiter should be investigated in detail in future research.
    }
    

    \subsection{Assumed Disk Model}\label{sec:Dis_ADM}
    In this study we adapt a relatively large disk in comparison to observed protoplanetary disks \citep[e.g.][]{Andrews+2010}.
    The planetesimal accretion rate is expected to change when considering different disk models.
    However, as long as Jupiter's formation pathway overlaps with the SSP, the second planetesimal accretion is expected to occur.
    To confirm that this is indeed the case, we performed additional simulations considering different disk models.
    The results are presented in Appendix~\ref{app:different_disk}. 
    Indeed, we find that a late phase of planetesimal accretion always occurs when proto-Jupiter enters the SSP. 
    We therefore conclude that the occurrence of second planetesimal accretion by proto-Jupiter is robust.
    
    In this study, we assumed that the planetesimals are homogeneously distributed.
    However, recent planetesimal formation models imply that planetesimals are distributed in a ring-like structure around ice lines because the solid-to-gas ratio is locally enhanced due to the pile-up of solid materials \citep{Armitage+2016,Drazkowska+2017,Hyodo+2021}.
    Ice lines of volatile materials such as $CO_2$ could lead to planetesimal formation at $\sim 10-15\AU$.
    In addition, the disk temperature evolution is important to consider in planetesimal formation models around ice-lines  \citep[e.g.,][]{Lichtenberg+2021}.
    We hope to investigate other planetesimal distributions and their time evolution in future studies. 
    Finally, it should be noted that the initial planetesimal distribution also affects the leftover distribution of small objects.
    By the end of the simulations in Case 2, more than $10M_{\oplus}$ of planetesimals remain in the region interior to Jupiter's orbit.
    These objects do not exist at present in the solar system, 
    \added{
    however,
    }
    the non-accreted planetesimals mainly come from distances $\lesssim10\AU$.
    \added{
    Therefore, if the planetesimals were formed around $10-15\AU$, the number of planetesimals shepherded into the region interior to Jupiter's orbit would be reduced.
    In addition, it is known that long-term dynamical evolution and gravitational interactions with other giant planets can lead to loss of planetesimals around Jupiter's orbit \citep{Dones+2004,OBrien+2007}.
    The orbital evolution of planetesimals after Jupiter's formation should be investigated in future work.
    }
    
\section{Conclusions}
\label{sec:conclusion}
    \begin{figure*}
      \begin{center}
        \includegraphics[width=160mm]{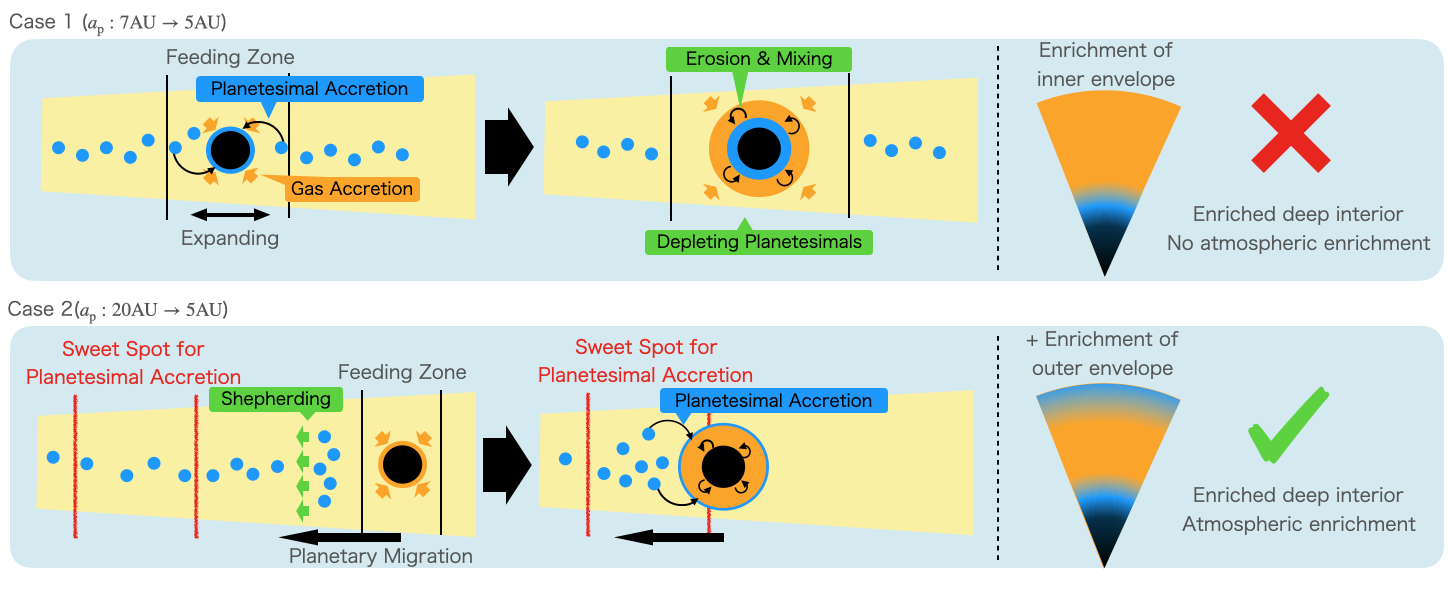}
        \caption{
        Sketch for planetesimal accretion and forming composition gradient in Jupiter envelope.
        The above panel shows the case when proto-Jupiter migrates from $7\AU$ to $5\AU$.
        In-falling gas covers the envelope enriched via planetesimal accretion and only inner region of Jupiter envelope is enriched with heavy elements.
        The bottom panel shows the case when proto-Jupiter migrates from $20\AU$ to $5\AU$.
        If the proto-planet reaches the sweet spot for accretion just before the end of gas accretion, planetesimals are deposited into the outer envelope.
        Several times enriched Jupiter's atmosphere can be formed if the size of outermost convective layer is so small as $\lesssim0.2 M_{\rm J}$.
        }
        \label{fig:sketch}
      \end{center}
    \end{figure*}
    
    We investigated Jupiter's origin focusing on the possibility of  planetesimal accretion towards the end of its formation.
    We considered two formation pathways: Case 1 where proto-Jupiter migrates from $\sim 7\AU$ to its current location and Case 2 where proto-Jupiter migrates from $\sim 20 \AU$.
    For Case 1, we find that the planetesimal accretion rate decreases with increasing planetary mass.
    Therefore for this case, Jupiter's outer envelope cannot be enriched with heavy elements.
    On the other hand, in Case 2, we find that a late  planetesimal accretion phase occurs before the end of Jupiter's  formation.
    This happens because proto-Jupiter enters the sweet spot for planetesimal accretion (see text for details), which leads to an enrichment of Jupiter's atmosphere. 
    
    The accreted heavy elements are expected to mix and redistribute in Jupiter's envelope during its long-term evolution.
    Assuming the mass of the uppermost layer of Jupiter's envelope $M_{\rm conv,top}$ and fully mixing of deposited heavy materials, we find that: 
    \begin{itemize}
        \item Jupiter's atmosphere is barely enriched in Case 1 regardless of the size of $M_{\rm conv,top}$.
        \item A relatively thin layer of $M_{\rm conv,top}\lesssim 0.2 M_{\rm J}$ in Case 2 is consistent with the observed metallicity of Jupiter's atmosphere.
    \end{itemize}
    The results of the two formation models we consider and their outcomes are summarised in the sketches of Fig.~\ref{fig:sketch}. 
    
    To conclude, we suggest that Jupiter's core was formed via pebble accretion and had migrated from $\sim 20$ AU to its current location followed by a late phase of planetesimal accretion that enriches its atmosphere with heavy elements. 
    In this scenario we infer an internal structure that is (at least qualitatively) consistent with interior models of Jupiter in which the outermost part of Jupiter's envelope is enriched with heavy elements by a factor of a few relative to the proto-solar composition.
    In our scenario, the atmospheric enrichment is a result of late planetesimal accretion and not due to convective mixing of heavy elements from the deep interior as suggested by other studies \citep[e.g.][]{Oberg+2019,Bosman+2019}.
    We conclude that Jupiter's atmosphere can be enriched with heavy elements if proto-Jupiter had migrated from $\sim 20$ AU to its current location.
    Further studies about the convection of Jupiter's envelope and the mixing of heavy materials would be used for the constraints of Jupiter's formation pathway. 


\acknowledgments

We acknowledge support from the Swiss National Science Foundation (SNSF) under grant \texttt{\detokenize{200020_188460}}. 

\newpage
\appendix
    \section{Planetary formation model}\label{app:model}
    \begin{figure}
      \begin{center}
        \includegraphics[width=80mm]{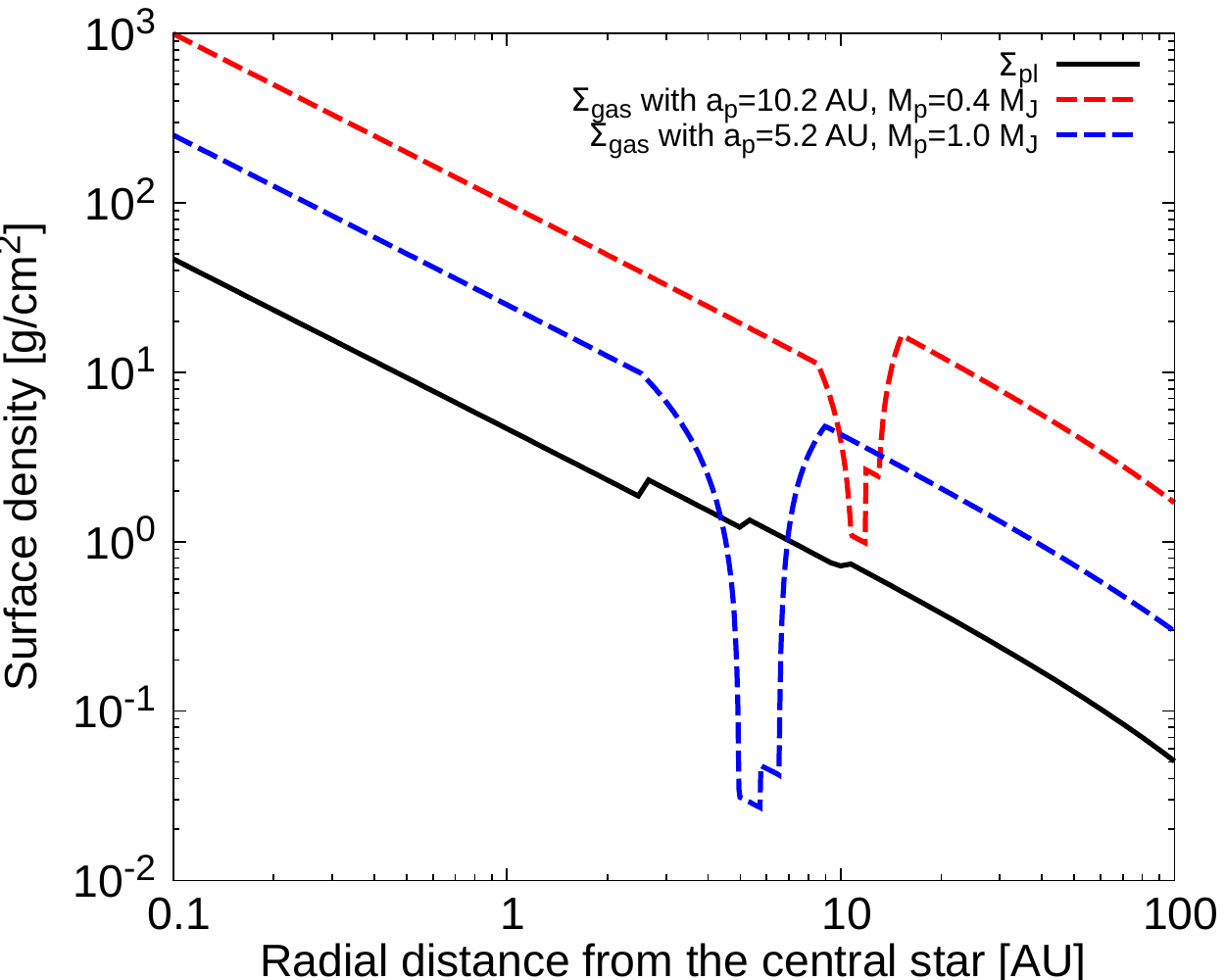}
        \caption{
        Disk profile used in our model.
        The solid line shows the planetesimal surface density profile.
        We adapt a solid-to-gas ratio used in \citet{Turrini+2021}.
        The dashed lines show the gas surface density profile when $M_{\rm p}=0.4 M_{\rm J}$ (red line) and $M_{\rm p}=1.0 M_{\rm J}$ (blue line) in Case 2.
        }
        \label{fig:Disk}
      \end{center}
    \end{figure}
    \begin{figure}
      \begin{center}
        \includegraphics[width=80mm]{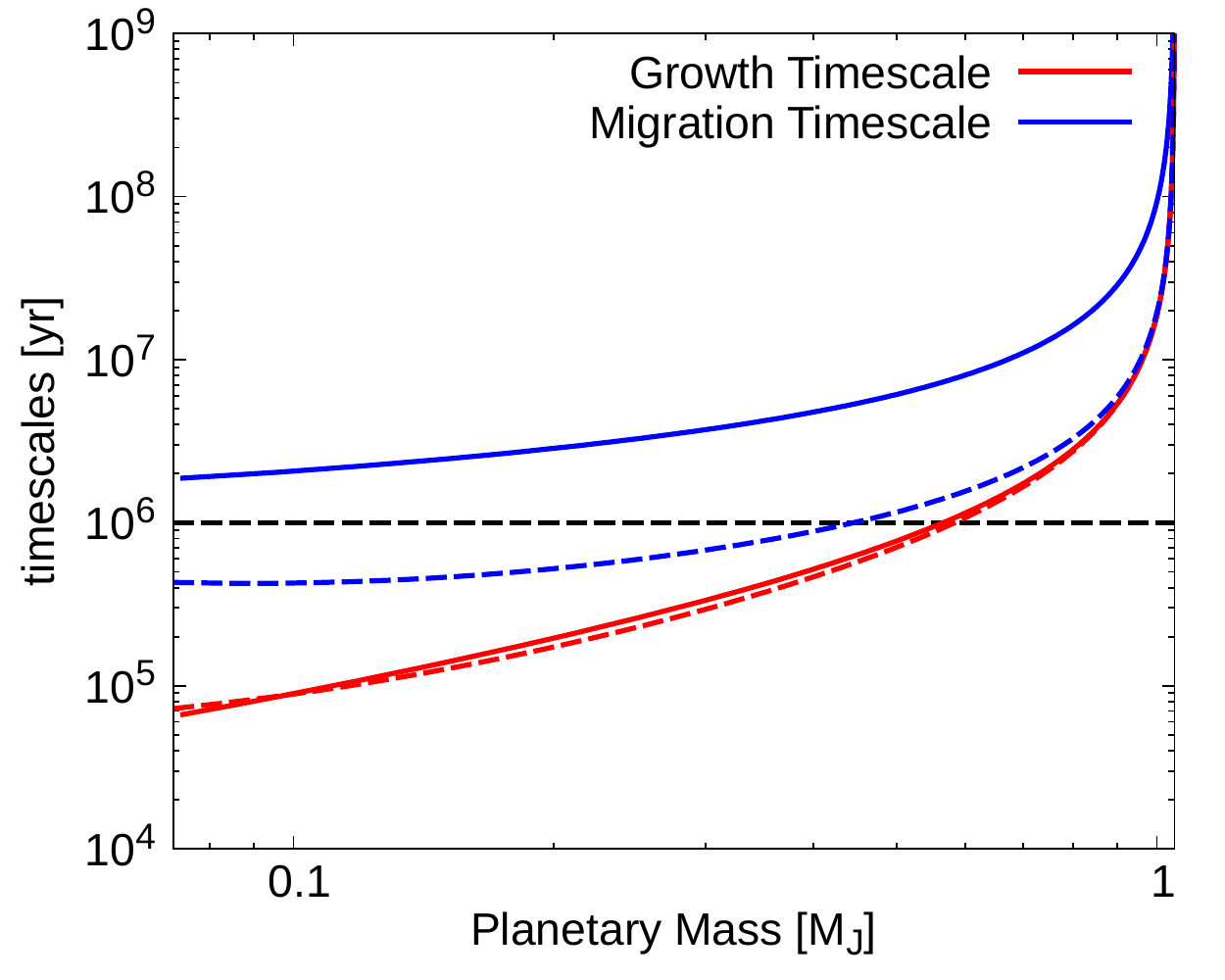}
        \caption{
        Evolution of timescales used in our model.
        Red and blue lines are gas accretion timescale $\tau_{\rm acc}$ and planetary migration timescale $\tau_{\rm tide}$, respectively.
        Solid and dashed lines are cases of $M_{\rm th}=10^{-2}$ and $10^{-3}$, respectively.
        The dotted black line shows the depletion timescale $\tau_{\rm dep}=1\times10^{6}\yr$.
        }
        \label{fig:mp_timescales}
      \end{center}
    \end{figure}
    
    We describe the formation model used in this study, which is based on the model by \citet{Tanaka+2020}.
    We adopt  
    \replaced{type II migration}{a planetary migration model}
    with a shallow gap empirically obtained by \citet{Kanagawa+2018}.
    The migration rate is given by: 
    \begin{align}
        \frac{d r_{\rm p}}{d t} = -2 c \frac{M_{\rm p}}{M_{\rm s}} \frac{{r_{\rm p}}^2 \Sigma_{\rm gap}}{M_{\rm s}} \left( \frac{h_{\rm p}}{r_{\rm p}} \right)^{-2} v_{\rm K,p}, \label{eq:migration_rate}
    \end{align}
    where $c$ is the constant set as $c=3$ in this study, $r_{\rm p}$ is the orbital radius of the planet, $\Sigma_{\rm gap}$ is the surface density of disk gas at the gap bottom, $h_{\rm p}$ is the disk gas scale height and $v_{\rm K,p}$ is the Kepler velocity of the planet.
    \added{
    During the late formation stage of gas giant planets, the gas accretion rate is regulated by the gas flow around the protoplaent rather than the cooling rate of the planetary envelope.
    In this case,
    }
    the gas accretion rate is given by   \citep{Tanigawa+2002}: 
    \begin{align}
        \frac{d M_{\rm p}}{d t} = D \Sigma_{\rm gap} \label{eq:gas_accretion_rate_Tanigawa}
    \end{align}
    with
    \begin{align}
        D = 0.29 \left( \frac{M_{\rm p}}{M_{\rm s}} \right)^{4/3} \left( \frac{h_{\rm p}}{r_{\rm p}} \right)^{-2} {r_{\rm p}}^2 \Omega_{\rm p} \label{eq:gas_accretion_rate}
    \end{align}
    where $\Omega_{\rm p}$ is the Kepler angular velocity of the planet.
    Using eq.~(\ref{eq:migration_rate}) and eq.~(\ref{eq:gas_accretion_rate_Tanigawa}), we can obtain eq.~(\ref{eq:timescale_fraction}) and $M_{\rm th}=10^{-2}$.
    In the Case 2, to account for the lower accretion rate, we artificially reduce the gas accretion rate by a factor of $\sim5$ and obtain $M_{\rm th}=10^{-3}$.
    
    Our baseline disk model is based on the self-similar solution for the surface density profile of disk gas \citep{Lynden-Bell+1974}.
    The mid-plane temperature of disk gas $T_{\rm disk}$ is given by: 
    \begin{align}
        T_{\rm disk} = 280 {\rm K} \left( \frac{r}{1\AU} \right)^{-1/2},
    \end{align}
    where $r$ is the radial distance from the central star.
    In this case, the disk gas viscosity $\nu=\alpha_{\rm vis} c_{\rm s} h_{\rm s}$, where $\alpha_{\rm vis}$ is the viscosity parameter \citep{Shakura+1973} and $c_{\rm s}$ is the sound speed of disk gas, is proportional to $r$ and the self-similar solution $\Sigma_{\rm SS}$ is given as: 
    \begin{align}
        \Sigma_{\rm SS} = \frac{M_{\rm tot,0}}{2 \pi {R_{\rm d}}^{2}} \left( \frac{r}{R_{\rm d}} \right)^{-1} T^{-3/2} \exp \left( -\frac{r}{T R_{\rm d}} \right), \label{eq:self-similar-solution}
    \end{align}
    with: 
    \begin{align}
        T &= 1 + \frac{t}{\tau_{\rm vis}}, \label{eq:normalized_time} \\
        \tau_{\rm vis} &= \frac{{R_{\rm d}}^2}{\nu_{\rm d}}, \label{eq:viscous_characteristic_timescale}
    \end{align}
    where $M_{\rm tot,0}$ is the disk total mass at $t=0$, $R_{\rm d}$ is a radial scaling length of protoplanetary disk, $\tau_{\rm vis}$ is the characteristic viscous timescale and $\nu_{\rm d}$ is a disk gas viscosity at $r=R_{\rm d}$.
    The surface density profile of disk gas is altered by the gap opening around the planet, the gas accretion onto the planet and the disk depletion.
    We include these effects and the surface density profile of disk gas $\Sigma_{\rm gas}$ is given by: 
    \begin{align}
        \Sigma_{\rm gas} = f_{\rm gap} f_{\rm acc} f_{\rm dep} \Sigma_{\rm SS}, \label{eq:Sigma_gas}
    \end{align}
    where $f_{\rm gap}$ is the gap opening factor, $f_{\rm acc}$ is the gas accretion factor, and $f_{\rm dep}$ is the disk depletion factor.
    For the gap opening factor, we adapt the empirically obtained model by \citet{Kanagawa+2017}.
    The gap structure changes with the radial distance from the planet $\Delta r=|r-r_{\rm p}|/r_{\rm p}$ and $f_{\rm gap}$ is written as a function of $\Delta r$ as
    \begin{align}
        f_{\rm gap} =
            \begin{cases}
                \displaystyle{            
                \frac{1}{1+0.04K} 
                } &{\rm for}~ \Delta r < \Delta R_1, \\
                \displaystyle{            
                4.0 {K^{\prime} }^{-1/4} \Delta r -0.32
                } &{\rm for}~\Delta R_1 < \Delta r < \Delta R_2 , \\
                \displaystyle{            
                1
                } &{\rm for}~ \Delta R_2 < \Delta r,
            \end{cases} \label{eq:gap_structure}.
    \end{align}
    with
    \begin{align}
        K           &= \left( \frac{M_{\rm p}}{M_{\rm s}} \right)^2 \left( \frac{h_{\rm p}}{r_{\rm p}} \right)^{-5} {\alpha_{\rm vis}}^{-1}, \\
        K^{\prime}  &= \left( \frac{M_{\rm p}}{M_{\rm s}} \right)^2 \left( \frac{h_{\rm p}}{r_{\rm p}} \right)^{-3} {\alpha_{\rm vis}}^{-1}, \\
        \Delta R_1  &= \left\{ \frac{1}{4(1+0.04K)} +0.08 \right\} {K^{\prime}}^{1/4}, \\
        \Delta R_2  &= 0.33 {K^{\prime}}^{1/4}.
    \end{align}
    In the disk region inner than the planet, disk surface density is reduced by the gas accretion onto the planet.
    When the gas accretion rate is given by eq.~(\ref{eq:gas_accretion_rate_Tanigawa}) and the gap structure is given by eq.~(\ref{eq:gap_structure}), $f_{\rm acc}$ is written as \citep{Tanaka+2020}
    \begin{align}
        f_{\rm acc} = 
            \begin{cases}
                1 &{\rm for}~r > r_{\rm p} \\
                \displaystyle{            
                \left\{ 1 + \frac{D}{3\pi \nu (1+0.04K)} \right\}^{-1}
                } &{\rm for}~r \leq r_{\rm p},
            \end{cases} \label{eq:sweet_spot}.
    \end{align}
    where $\nu$ is the disk gas viscosity given as $\nu=\alpha_{\rm vis} c_{\rm s} h_{\rm s}$.
    To account for the disk depletion process, such as photo-evaporation or disk wind, we set the disk depletion factor $f_{\rm dep}$ as
    \begin{align}
        f_{\rm dep} = \exp \left( - \frac{t}{\tau_{\rm dep}} \right).
    \end{align}
    The surface density of disk gas at the gap bottom, which is used for the gas accretion rate and migration rate, is obtained as $\Sigma_{\rm gap}=\Sigma_{\rm gas} (r=r_{\rm p})$ using eq.~(\ref{eq:Sigma_gas}).
    
    In our model, we set $\alpha_{\rm vis}=10^{-3}$, $M_{\rm disk,0}=0.1M_{\odot}$ and $R_{\rm disk}=200\AU$, respectively.
    The final orbital position of proto-Jupiter at the $a_{\rm p}-M_{\rm p}$ plane is determined by the core formation time $t_0$ \citep[e.g.][]{Tanaka+2020}.
    We find that Jupiter stops at its current location due to the disk depletion if the core formed with $t_0=1.3\times10^6 \yr$ and $t_0=0.9\times10^6 \yr$ for Case 1 and Case 2, respectively.
    We continue the orbital integration for $1\times10^7 \yr$.
    By the end of the simulation, we find that Jupiter does not grow further and that planetesimal accretion is negligible at that stage. 
    
    \replaced{For simplicity}{For the planetesimal disk}, we assume that the planetesimal distribution follows the density profile of the gaseous disk at $t=0$ and adopt the solid-to-gas ratio used in \citet{Turrini+2021}. 
    To speed up the numerical simulation, we adopt super-particles used in \citet{Shibata+2021}.
    \added{
    We use $12000$ super-particles in Case 1 and $24000$ super-particles in Case 2.
    Super-particles are distributed from $3.9 \AU$ to $8.4 \AU$ in Case 1 and from $3.9 \AU$ to $23\AU$ in Case 2.
    The initial eccentricity $e$ and inclination $i$ of planetesimals are given by the Rayleigh distribution and we set $\left< e^2 \right>^{1/2}=2\left< i^2 \right>^{1/2}=10^{-3}$.
    The other orbital angles are distributed uniformly.
    }
    
    Figure~\ref{fig:Disk} shows the disk model used in this letter.
    The solid and dashed lines are the surface density of solid materials (or planetesimals) and gas, respectively.
    Figure~\ref{fig:mp_timescales} shows the evolution of gas accretion timescale $\tau_{\rm acc}$ (red) and planetary migration timescale $\tau_{\rm tide}$ (blue) in our model.
    The solid and dashed lines are cases of $M_{\rm th}=10^{-2}$ and $10^{-3}$, respectively.
    Both timescales rapidly increase around $M_{\rm p}\sim M_{\rm J}$ due to the exponential decay of disk gas.
    We stop our simulation at $t-t_0=1\times10^{7}\yr$.
    Even if we proceeded the simulation farther, proto-Jupiter would not grow and migrate anymore.
    When $M_{\rm th}=10^{-2}$, the fraction of timescales keeps a large value of $\tau_{\rm tide}/\tau_{\rm acc}\gg1$.
    \added{
    In this case, the planetary  mass reaches Jupiter's mass before the disk dissipation because $\tau_{\rm acc}$ is smaller/comparable to the disk's  depletion timescale $\tau_{\rm dep}$.
    However, the planet barely migrates because $\tau_{\rm acc}$ is always longer than $\tau_{\rm dep}$.
    }
    On the other hand, when $M_{\rm th}=10^{-3}$, the fraction of timescales decreases to $\sim1$.
    \added{
    Both timescales are smaller than or comparable to $\tau_{\rm dep}$, so  proto-Jupiter can migrate over significant distances ($\sim$ $10\AU$) before the disk is dissipated. 
    }

    \added{
    In our simulations, planetesimal capture is considered to occur once a planetesimal enter proto-Jupiter's envelope.
    The radius of proto-Jupiter is given by
    \begin{align}
        R_{\rm p} = \left( \frac{3 M_{\rm p}}{4 \pi \rho_{\rm p}} \right)^{1/3} \label{eq:capture_radius},
    \end{align}
    where $\rho_{\rm p}$ is the mean density.
    During the detached phase, the planetary envelope is slightly expanded due to the gas accretion \citep{Valletta+2021}.
    To include the effect, we set $\rho_{\rm p}=0.125 \g/\cm^3$.
    $R_{\rm p}$ in our model is always smaller than that obtained by \citet{Valletta+2021} by a factor of a few, and it is knows that the capture radius is larger for smaller planetesimals \citep[e.g.][]{Inaba+2003}.
    We do not expect significant  differences if a more detailed calculation of the capture radius is considered. 
    This could be explored in detail in future research.
    }
    Table \ref{tb:model_settings} shows the parameters used in this study.
    
    \begin{table}
    	\centering
    	\begin{tabular}{ llllll } 
    		\hline
    		\hline
    		$M_{\rm s}$ 		& Mass of central star	                        & $1.0 \Msun$ &	\\
    		$M_{\rm disk,0}$    & Initial mass of protoplanetary disk           & $0.1 \Msun$ &	\\
    		$R_{\rm disk}$      & Typical size of protoplanetary disk           & $200 \AU$ &	\\
    		$\alpha$            & Disk viscosity parameter                      & $1\times10^{-3}$ \\
    		$\tau_{\rm dep}$ 	& Disk depletion timescale                      & $1\times10^6 \yr$ & \\
    		$M_{\rm p,0}$ 		& Initial mass of protoplanet	                & $6\times10^{-5} \Msun$ & \\
    		\hline
    		\hline
    		                    &  & in Case 1 & in Case 2 \\
    		\hline
    		$a_{\rm p,0}$ 		& Initial semi-major axis of protoplanet        & $6.9 \AU$ & $18.3 \AU$  \\
    		$t_{\rm 0}$ 		& Formation time of planetary core		        & $1.3\times10^6 \yr$ & $0.9\times10^6 \yr$ \\
    		\hline
    	\end{tabular}
    	\caption{
    	Parameters used in our simulations.
        }
        \label{tb:model_settings}
    \end{table}
    
    \section{The effect of the assumed disk's profile}\label{app:different_disk}
    \begin{figure}
      \begin{center}
        \includegraphics[width=160mm]{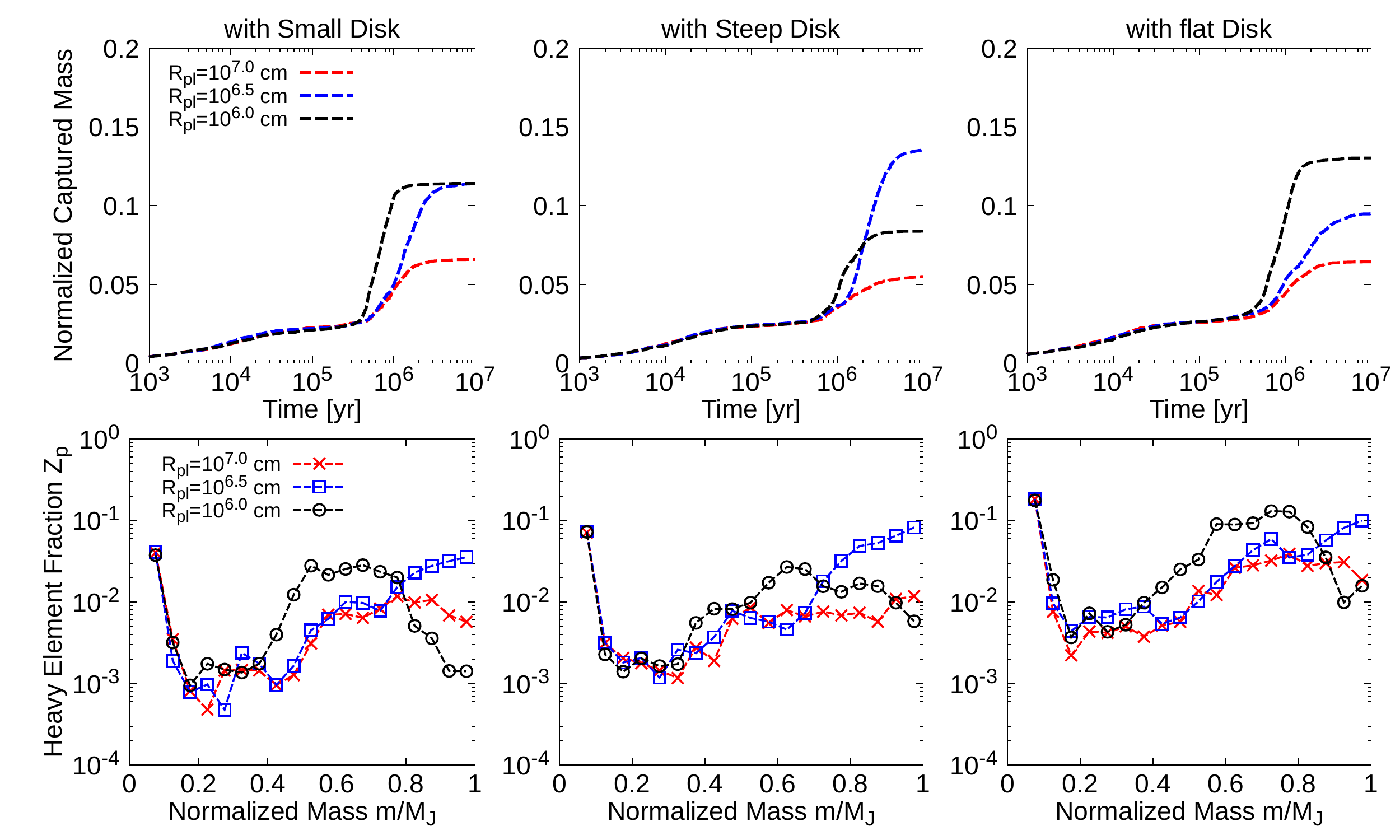}
        \caption{
        Same as Fig.~\ref{fig:Result} but using  small disk model (left),  steep disk model (middle) and  flat disk model (right).
        In the upper panels, we normalise the cumulative captured mass using the total solid mass used in each simulations.
        We adapt the evolution pathways of Case 2.
        }
        \label{fig:time_Mcap_disk}
      \end{center}
    \end{figure}
    In Sec.~\ref{sec:results}, we adapt a self-similar solution for a protoplanetary disk with $M_{\rm disk,0}=0.1M_{\odot}$ and $R_{\rm disk}=200\AU$. However, the structure and the size of protoplanetary disks are not well-determined. 
    Here, we show the results when assuming three different disk models.
    The first one is a small disk model where we adapt the self-similar solution but with $M_{\rm disk,0}=0.03M_{\odot}$ and $R_{\rm disk}=50\AU$.
    The second and third disk models are a steep disk model and a flat disk model where we adapt a simple disk profile given by: 
    \begin{align}
        \Sigma_{\rm Simple} = \Sigma_{0} \left( \frac{r}{5.2 \AU} \right)^{-\alpha_{\rm disk}}
    \end{align}
    with $\alpha_{\rm disk}=3/2$ for the steep disk and $\alpha_{\rm disk}=1/2$ for the flat disk.
    $\Sigma_{0}$ is set as $300\g/\cm^2$ and the gap opening factor, the gas accretion factor and the disk depletion factor are adapted in the same way as eq.~(\ref{eq:Sigma_gas}).
    Due to the different distribution of the disk gas, the evolution is different from the original model.
    However, the evolution pathways on the $a_{\rm p}-M_{\rm p}$ plane is similar to the baseline model because the timescale fractions are independent of the disk's profile (see Eq.~\ref{eq:timescale_fraction}).
    
    Fig.~\ref{fig:time_Mcap_disk} shows the results using the various disk models.
    The total captured mass of planetesimals and the timing of planetesimal accretion is similar to the baseline disk model.
    This is because the location of the SSP is nearly independent of the disk surface density profile \citep{Shibata+2021}.
    Even if the gaseous disk had a different density distribution, the SSP would locate around Jupiter's  orbit.
    Therefore we conclude that the occurrence of a second planetesimal accretion phase where  heavy elements are deposited into the upper envelope of proto-Jupiter is robust and does not depend on the assumed disk model.

    \section{Planetesimal collisions}\label{sec:Dis_PCL}
    In our simulation, we adopt test particles for planetesimals and neglect collisions between planetesimals.
    As pointed by \citet{Batygin+2015b} and \citet{Shibata+2021}, planetesimal collisions cloud be important during the shepherding process.
    We find that more than 10 M$_{\oplus}$ of planetesimals are shepherded by the mean motion resonances
    \added{
    before proto-Jupiter enters the SSP.
    }
    These planetesimals could collide with each other as the planet migrates inwards.
    Once collisional cascade begins, the planetesimal size distribution can change and therefore affecting (i.e., reducing) the planetesimal accretion rate.
    
    \added{
    The collision timescale depends on the total mass of the planetesimals shepherded by the mean motion resonances.
    At the same time, the total mass shepherded by the mean motion resonances changes with the planetesimal distribution.
    If the planetesimal disk is formed around the ice lines and has a ring-like distribution around $10-15\AU$ which is different from the uniform distribution used in our simulations, the total mass of planetesimals shepherded by the mean motion resonances would be smaller than $10 M_{\oplus}$.
    In that case the collision timescale would be longer than the migration timescale and the second planetesimal accretion phase would start before the initiation of the collisional cascade (see \citet{Shibata+2021} for further details).
    It is clear that the planetesimal distribution plays a key role in this process and in some cases could prevent the initiation of collisional cascade.
    }
    
    \section{Effect of other planets}
    Our simulations focused on the interaction between a migrating planet and the surrounding planetesimals and do not include the existence of other protoplanets.
    The gravitational perturbations from other protoplanets, however, could affect the location of the SSP \citep{Shibata+2020}.
    In addition, the migration of other planets can change the distribution of planetesimals, and even contribute the further planetesimal formation  \citep{Shibaike+2020}.
    It is therefore clear that future studies should  investigate formation pathways accounting for the growth of all the outer planets and their mutual interactions.
    Since the atmospheres of all the outer planets in the solar system are measured to be enriched with heavy materials \citep[e.g.][]{Atreya+2020} it is desirable to investigate planetesimal accretion mechanisms for all four planets.

\bibliographystyle{aasjournal} %
\bibliography{refs} %

\listofchanges

\end{document}